\documentclass{llncs}
\usepackage{epic,eepic,amsmath,latexsym,amssymb,color}
\usepackage{ifthen,graphics,epsfig}
\usepackage[english]{babel}
\usepackage{times}

\newlength {\squarewidth}

\newcommand{\toto}{xxx}
\newenvironment{proofT}{\noindent{\bf Proof }}
{\hspace*{\fill}$\Box_{Theorem~\ref{\toto}}$\par\vspace{3mm}}
\newenvironment{proofL}{\noindent{\bf Proof }}
{\hspace*{\fill}$\Box_{Lemma~\ref{\toto}}$\par\vspace{3mm}}

\newcommand{\ggcd}{{\sf{gcd}}}
\newcommand{\ppmod}{{~\sf{mod~}}}
\newcommand{\order}{{\sf{order}}}
\newcommand{\REG}{\mathit{REG}}

\def\ffalse{\mathit{false}}
\def\ttrue{\mathit{true}}

\def\won{\mathit{won}}

\def\ssum{\mathit{sum}}

\def\old{\tt{old}}
\def\new{\tt{new}}
\def\compareandswap{{\sf{compare\&swap}}}

\def\myview{\mathit{myview}}
\def\round{\mathit{round}}
\def\counter{\mathit{counter}}

\def\competitors{\mathit{competitors}}

\def\ffalse{\tt{false}}
\def\ttrue{\tt{true}}
\def\gcd{{\sf gcd}}
\def\return{{{\sf return}}}

\def\elect{{\sf elect}}

\newcommand{\pp}{\mathit{R}}

\newcommand{\MMR}[1]{{#1}}

\date{}

{\title{\bf{Election in Fully Anonymous Shared Memory Systems:\\
        Tight Space Bounds and Algorithms}}}

\author{Damien Imbs${^\dag}$, Michel Raynal$^{\star}$,
        Gadi Taubenfeld${^\circ}$}
\institute{
${^\dag}$LIS, Aix-Marseille University \&CNRS \& Univ. Toulon,  France\\
$^{\star}$Univ Rennes IRISA, Inria, CNRS, France \\
        ${^\circ}$Reichman University, Herzliya, Israel}

\begin{document}

\maketitle

\begin{abstract}
This article addresses election in fully anonymous systems made up of  $n$
asynchronous processes that communicate through atomic read-write registers
or atomic read-modify-write registers.  Given
an integer $d\in\{1,\dots, n-1\}$, two elections
problems are considered: {\it $d$-election} (at least one and at most
$d$ processes are elected) and {\it exact $d$-election} (exactly $d$
processes are elected).  Full anonymity means that both the processes
and the shared registers are anonymous. Memory anonymity means that
the processes may disagree on the names of the
shared registers. That is, the same register name $A$ can denote different
registers for different processes, and the register name $A$ used by a
process and the register name $B$ used by another process can address
the same shared register.
Let $n$ be the number of processes, $m$ the number of atomic
read-modify-write registers, and
let $M(n,d) =\{k~:~ \forall ~\ell : 1 < \ell\leq n\ : ~\gcd(\ell,k)\leq d\}$.
The following results are presented
for solving election in such an adversarial full anonymity context.
\begin{itemize}
\item
It is possible to solve $d$-election when participation is not required
if and only if $m\in M(n,d)$.
\item
It is possible to solve exact $d$-election when participation is required
if and only if $\ggcd(m,n)$ divides $d$.
\item
It is possible to solve $d$-election when participation is required
if and only if $\ggcd(m,n)\leq d$.
\item
  Neither  $d$-election nor exact $d$-election (be participation required
  or not)  can be solved when
the processes communicate through read-write registers only.
\end{itemize}

\end{abstract}


\section{Introduction}

\subsection{Leader election}
{\it Leader election} is a classic basic problem encountered when
processes cooperate and coordinate to solve higher-level
distributed computing problems.  It consists in designing an algorithm
selecting one and only one process from the set of cooperating
processes.  In classical systems where the processes have distinct
identities, leader election algorithms usually amount to electing the
process with the smallest (or highest) identity. Many textbooks
describe such algorithms (e.g., \cite{AW04,KS08,R13,T06}).

This article considers two natural generalizations of the election
problem in the presence of both process and memory anonymity, where
communication is through shared registers.  The first one is {\it
  $d$-election} in which at least one and at most $d$ processes are
elected.  The second one is \emph{exact $d$-election} in which exactly
$d$ (different) processes are elected.

\subsection{System models}

\paragraph{Process anonymity}
Process anonymity means that the processes have no identity, have the same
code, and have the same initialization of their local variables.  Hence, in
a process anonymous system, it is impossible to distinguish a process
from another process.

%
Pioneering work on process anonymity in
message-passing systems was presented in~\cite{A80}.
Process anonymity has been studied for a long time in
asynchronous shared memory systems (e.g., \cite{AGM02}).  It has been
more recently addressed in the context of crash-prone asynchronous
shared memory systems (e.g., \cite{BRS18,GR07}).

Assuming a system made up of $n$ anonymous asynchronous processes, we
use the notation $p_1$, ..., $p_n$ to distinguish the processes. The
subscript $i\in\{1,\cdots,n\}$ will also be used to identify the local
variables of $p_i$ (identified with names written with lower case
letters).

\paragraph{Shared registers}
The processes communicate through a shared memory made up of $m$
atomic registers~\cite{L86} (identified with names written with upper
case letters).  Hence, the shared memory appears to the processes as an
array of registers denoted $R[1..m]$.  {\it Atomic} means that
the operations on a register appear as if they have been executed
sequentially, each appearing between its start event and its end
event~\cite{HW90}. Moreover, for any $x$, a read of $R[x]$ returns the
last value previously written in $R[x]$, where {\it last} refers to
the previous total order on the operations on $R[x]$.  (In case $R[x]$
has not been written, the read returns its initial value.)  Two
communication models are considered in the article.
\begin{itemize}
\item Read-write (RW) model.
  This is the basic model in which a register  $R[x]$ can be
  accessed only by a read or a write operation (as the
  cells of a Turing machine).
\item Read-modify-write (RMW) model.
  This model is  the RW model enriched with a conditional write operation.
  This conditional write operation atomically  reads the register and
  (according to the value read) possibly modifies it.
  This  conditional write, denoted $\compareandswap(R[x],\old,\new)$,
  has three parameters, a shared register, and  two values.
  It returns a Boolean value.  If $R[x]=\old$, it assigns the value $\new$
  to  $R[x]$ and returns $\ttrue$.   If $R[x]\neq  \old$,  $R[x]$
  is not modified and the operation returns  $\ffalse$.
 An  invocation of $\compareandswap(R[x],\old,\new)$ that returns
$\ttrue$ is {\it successful}.
\end{itemize}

\paragraph{Memory anonymity}
\label{def:anonymous-memory}
The notion of an anonymous memory has been introduced in~\cite{T17}.
In a non-anonymous memory,  the address $R[x]$ denotes the same register
whatever the process that invokes $R[x]$: there is an a priori agreement
on the name of each register.
In an anonymous memory, there is no such agreement on the names
of the shared  registers.
While the name $R[x]$ used by a given process $p_i$
always denotes the very same register,
the same name $R[x]$ used by different processes $p_i$ $p_j$, $p_k$ ..,
may refer to different registers. More precisely, an anonymous memory
system is such that:
\begin{itemize}
\item For each  process $p_i$, an adversary defined a permutation
  $f_i()$ over the set $\{1,2,...,n\}$ such that, when $p_i$ uses the
  name $R[x]$, it actually accesses $R[f_i(x)]$,
\item No process knows the permutations,
\item All the registers are initialized to the same default value.
\end{itemize}

In an anonymous memory system, ALL the registers are anonymous.
  Moreover, the  size of the anonymous memory is not under
  the control of the programmer, and it is imposed on her/him.
  As shown in~\cite{AIRTW19,GIRT20-podc} (for non-anonymous processes and
anonymous memory), and in this article (for fully anonymous systems)
the size of the anonymous memory is a  crucial parameter when one has to
characterize  the pairs $\langle n,m\rangle$ for which election can
be solved in fully anonymous $n$-process systems.

\paragraph{Process participation}
As in previous works on election in
anonymous or non-anonymous memory systems~\cite{GIRT20-podc,SP89},
  this article
  considers two types of assumptions on the behavior of the processes:
  (1) algorithms that require the participation of all the
  processes to compete to be leaders, and (2) algorithms that do not
  (i.e., an arbitrary  subset of processes may participate
   but not necessarily all the processes).

\paragraph{Symmetric algorithm}
Considering a system in which the processes have distinct identifiers,
a  {\it symmetric} algorithm is  an algorithm where the processes
can only  compare their identities  with equality~\cite{SP89}.
So there is no notion of smaller/greater on process identities,
and those cannot be used to index entries of arrays, etc.
This notion of symmetry associated with  process identities
is the ``last step'' before their anonymity.
In this article, we will consider algorithms in which the processes
(and memories) are anonymous
but will also mention symmetric algorithms.

\subsection{Related works on anonymous memories}
Since its introduction, several problems have been
addressed in the context of memory anonymity: mutual exclusion,
election, consensus, set-agreement and renaming.
We discuss below work on the first two
problems that are more related to our work.
\paragraph{Mutual exclusion}
First, we observe that no shared memory-based mutual exclusion
algorithm requires the participation of the processes.
Let $M(n) = \{k~:~ \forall ~\ell : 1 < \ell\leq n\ : ~\gcd(\ell,k)=1\}$
(all the integers $2, ..., n$ are relatively prime with $k$).
The following results have been recently established.
\begin{itemize}
\item There is a deadlock-free symmetric mutual exclusion algorithm in
  the RMW (resp. RW) model made up of $m$ anonymous registers if and
  only if $m\in M(n)$ (resp. $m\in M(n)\setminus \{1\}$)
  \cite{AIRTW19}.
\item There is a deadlock-free  mutual exclusion algorithm
  in the process anonymous and memory anonymous RMW model made
  up of $m$ registers if and only if $m\in M(n)$. Moreover, there is no
  such algorithm in the fully anonymous RW communication model~\cite{RT20a}.
\end{itemize}
The conditions relating $m$ an $n$ can be seen as the seed needed to
break symmetry despite anonymous memory, and symmetric or anonymous
processes, thereby allowing mutual exclusion and election to be
solved.  A single leader election can be considered one-shot mutual
exclusion, where the first process to enter its critical section is
elected.

\paragraph{Election in the symmetric model}
Considering the symmetric process model
in which all the processes (unlike in the mutual exclusion problem)
are required to participate. The following results are
presented in~\cite{GIRT20-podc} for such a model.
\begin{itemize}
\item
  There is a $d$-election symmetric  algorithm
  in the  memory anonymous RW and  RMW communication models made
  up of $m$ registers  if and only if $\ggcd(m,n)\leq d$.
\item
  There is an exact $d$-election symmetric  algorithm
  in the memory anonymous RW and  RMW communication  models made
  up of $m$ registers if and only if $\ggcd(m,n)$ divides $d$.
\end{itemize}
We emphasize that the above results for $d$-election assume that the
processes are symmetric and not that they are anonymous, as done in
this article.  Finally, fully anonymous agreement problems are
investigated in \cite{RT20b}.

  \paragraph{Remark}
  While addressing a different problem in a different context,
  it is worth mentioning  the work presented in ~\cite{FIPS13}
  that addresses the exploration of
  an $m$-size anonymous not-oriented ring by a team of  $n$
  identical, oblivious,  asynchronous mobile robots
  that can view the environment but cannot communicate.
  Among other results, the authors have shown that there
  are initial placements of the robots for which gathering
  is impossible when $n$ and $m$  are not co-prime, i.e., when
  $\gcd(n,m)\neq 1$.
  They also show that the problem is always solvable
 for $\gcd(n,m)= 1$  when $n\geq 17$.

\subsection{Motivation and content}

\paragraph{Motivation}
\MMR{
The main motivation of this work is theoretical.
It investigates a fundamental symmetry-breaking
problem (election) in the worst  adversarial context, namely
asynchronous and fully anonymous systems.
Knowing what is possible/impossible, stating
computability and complexity lower/upper bounds
are at the core of algorithmics~\cite{A80,HF12}, and trying to find
solutions ``as simple as possible'' is a key if one wants to
be able to master the complexity of  future
applications~\cite{AZ10,D80}.  This article aims to increase our
knowledge of what can/cannot be done in the full anonymity
context, providing associated
necessary and sufficient conditions which enrich our knowledge  on the
system assumptions under which  fundamental problems such as  election
 can be solved.}\footnote{On an application-oriented side,
it has been shown in~\cite{NB11,NB15,RTB21} that the process of
genome-wide epigenetic modifications (which allows cells to utilize
the DNA) can be modeled as a fully anonymous shared memory system
where, in addition to the shared memory, also the processes (that is
proteins modifiers) are anonymous. Hence fully anonymous systems can
be useful in the context of biologically-inspired distributed
systems~\cite{NB15,RTB21}.}

When one has to solve a symmetry-breaking problem, the main issue
  consists in finding the ``as weak as possible'' initial seed from which
  the initial system symmetry can be broken.
  So, considering FULL anonymity, this article complements previously known
  results on anonymous systems (which were on non-anonymous processes
  and anonymous memory~\cite{AIRTW19,GIRT20-podc}).  In all cases, the
  seed that allows breaking the very strong (adversary) symmetry
  context defined by FULL anonymity
  is captured by necessary and sufficient conditions
  relating the number of anonymous processes and the size of the
  anonymous memory.

\paragraph{Content of the article}
Let $m, n$ and $d$ be the number of registers, the number of processes and
the number of leaders,
respectively, and (as previously defined)
$M(n,d) =\{k~:~ \forall ~\ell : 1 < \ell\leq
n\ : ~\gcd(\ell,k)\leq d\}$.  Table \ref{table:global-view} summarizes
the four main results.
%
\begin{table}[ht]
\begin{center}
\renewcommand{\baselinestretch}{1}
\small
\begin{tabular}{|c||c|c|c|c|c|c|}
\hline
Problem & Register & Participation & Necessary \& sufficient
                                          & Nece- & suff- & Section\\
& type &&condition on $\langle m,n,d\rangle$&ssary& icient &\\
\hline
\hline
$d$-election & RMW & not required  & $m\in M(n,d)$
& Thm \ref{theorem-imposs-d-election} & Thm \ref{theorem-algorithm-1}& \ref{sec-d-election}\\
&&&&& Algo 1& \\
\hline
Exact $d$-election & RMW & required & $\ggcd(m,n)$ divides $d$
& Follows &Thm \ref{theorem-algorithm-2} &\ref{sec-exact-d-election}\\
&&&&from \cite{GIRT20-podc}& Algo 2&\\
\hline $d$-election & RMW & required & $\gcd(m,n)\leq d$& Follows &
Thm~\ref{theorem-algorithm-3}&\ref{sec-d-election-required}\\ &&&&from
\cite{GIRT20-podc}& &\\
\hline
$d$-election and & RW & required or & Impossible
&&&\ref{sec-RW-impossibility}\\
Exact $d$-election && not required &
Corollary~\ref{coro--imposs-RW-model} &&&\\
\hline
\end{tabular}
\end{center}
\caption{Election in the fully anonymous shared memory systems.}
\label{table:global-view}
\end{table}

\begin{enumerate}
\item
  A $d$-election algorithm in the RMW communication model, which does
  not require participation of all the processes.  It is also shown
  that the condition $m\in M(n,d)$ is necessary and sufficient for
  such an algorithm.  Notice that $M(n,1)$ is the set $M(n)$ that
  appears in the results for fully anonymous mutual exclusion
  discussed earlier.
\item
  An exact $d$-election algorithm for the RMW communication model in
  which all the processes are required to participate. It is also
  shown that the necessary and sufficient condition for such an
  algorithm is  $\ggcd(m,n)$ divides $d$.
\item
  A $d$-election algorithm (which is based on the previous result) for
  the RMW communication model in which all the processes are required
  to participate. It is also shown that $\ggcd(m,n)\leq d$ is a
  necessary and sufficient  condition for such an algorithm. (The
  short algorithm appears in the proof of
  Theorem~\ref{theorem-algorithm-3}.)\footnote{Both the algorithms
    described in the paper are simple. Their early versions were far
    from being simple, and simplicity is a first class property.
    As said by Y. Perlis (the recipient of first Turing Award)
    ``{\it Simplicity does not precede complexity,
      but follows it}''~\cite{P82}.}
\item
  An impossibility result that regardless whether participation is
  required or not, there is neither $d$-election nor exact
  $d$-election algorithm in the anonymous RW communication model.
\end{enumerate}
Let us notice that, due to the very nature of the anonymous process model,
no process can know the ``identity''  of  elected processes.
So, at the end of an election algorithm in the anonymous process model,
a process only knows if it is or not a  leader.

We point out that the leader election problem has several variants,
and the most general one, where a process
only knows if it is or not a  leader is a very common variant
\cite{AW04,R13b,T06}.


\section{\emph{d}-Election in the RMW Model \\
  Where Participation is Not Required}
\label{sec-d-election}

Throughout this section, it is assumed that communication is through
RMW anonymous registers and that the processes are not required to
participate.

\subsection{A necessary condition for \emph{d}-election}
\label{imposs-d-election-anonymous-memory}
In this subsection, it is further assumed that processes have
identities that can only be compared (symmetry constraint).  As they
are weaker models, it follows that the necessary condition proved
below still holds in RMW model where both the processes and the memory
are anonymous, and in the model where communication is through
anonymous RW registers.

\begin{theorem}
\label{theorem-imposs-d-election}
There is no symmetric $d$-election algorithm in the {\em RMW}
communication model for $n\geq 2$ processes using $m$ anonymous
registers if $m\notin M(n,d)$.
\end{theorem}

\begin{proofT}
Let $k$ be an arbitrary positive number such that $1\leq k \leq
n$. Below we examine what must be the relation between $k$, $m$ and
$d$, when assuming the existence of a symmetric $d$-election algorithm
for $n$ processes using $m\geq 1$ anonymous RMW registers.  To
simplify the modulo notation, the processes are denoted $p_0$, ..., $p_{n-1}$.

Let $\ggcd(m,k)= \delta$, for some positive number $\delta$.  We will
construct a run in which exactly $k$ processes participate.  Let us
partition these $k$ processes into $\delta \geq 1$ disjoint sets,
denoted $P_0,...,P_{\delta -1}$, such that there are exactly $k/
\delta$ processes in each set.  This partitioning  is achieved by
assigning process $p_i$ (where $i\in \{0,...,k-1\}$) to the set
$P_{i\ppmod{\delta}}$.  For example, when $k=6$ and $\delta =3$, $P_0
= \{p_0,p_3\}$, $P_1 = \{p_1,p_4\}$, and $P_2 = \{p_2,p_5\}$
(top of Figure~\ref{fig:proof-theorem-RMW-necessary}).  Such a
division is possible since, by definition, $\ggcd(m,k)=~\delta$.

\begin{figure}[ht]
\centering{
\includegraphics[width=0.45\textwidth]{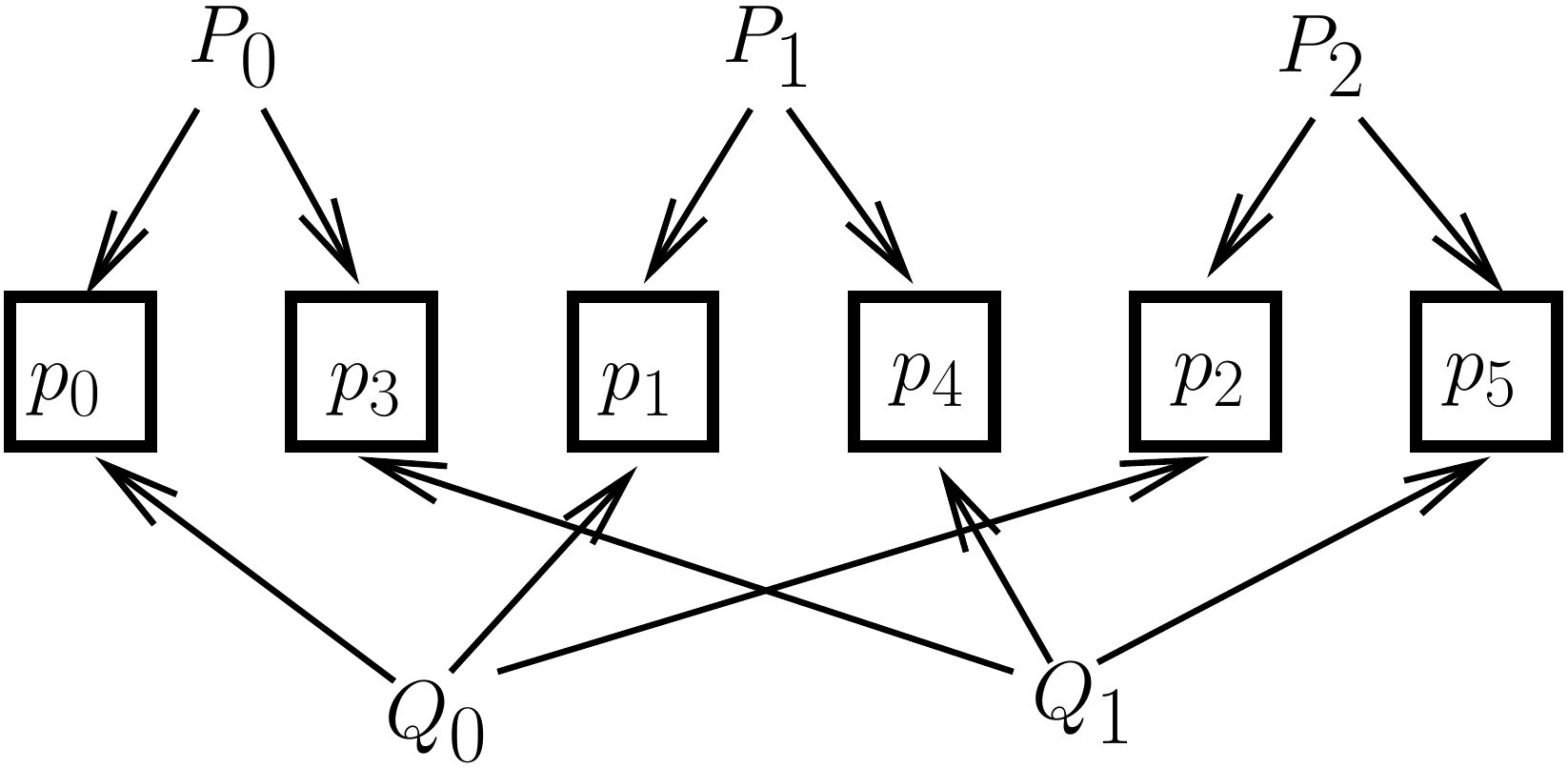}
\caption{Illustration of the runs for $k=6$ and $\delta=3$}
\label{fig:proof-theorem-RMW-necessary}
}
\end{figure}

Let us arrange the $m$ registers on a ring with $m$ nodes where each
register is placed on a different node.  To each one of the $\delta$
sets of processes $P_i$ (where $i\in \{0,...,\delta -1\}$), let us
assign an initial register (namely, the first register that each
process in that set  accesses) such that for every two sets $P_i$
and its ring successor $P_{(i+1) \ppmod{\delta}}$ the distance between
their assigned initial registers is exactly $\delta$ when walking on
the ring in a clockwise direction. This is possible since $\ggcd(m,k)=
\delta$.

The lack of global names for the RMW anonymous registers allows us
to assign, for
each one of the $k$ processes,   an initial register and an ordering
which determines how the process scans the registers.
An execution in which the $k$ processes are running in
\emph{lock-steps}, is an execution where we let each process take one
step (in the order $p_0,...,p_{k-1}$), and then let each process take
another step, and so on. For a given $d$-election algorithm $A$, let
us call this execution, in which the processes run in lock-steps,
$\rho_A$. For simplicity, we will omit the subscript $A$ and simply
write $\rho$.

For process $p_i$ and integer $j$, let $\order(p_i, j)$ denotes the
$j^{th}$ new (i.e., not yet assigned) register that $p_i$ accesses
during the execution $\rho$, and assume that we arrange that
$\order(p_i, j)$ is the register whose distance from $p_i$'s initial
register is exactly $(j-1)$, when walking on the ring in a clockwise
direction.

We notice that $\order(p_i, 1)$ is $p_i$'s initial register,
$\order(p_i, 2)$ is the next new register that $p_i$ accesses and so
on.  That is, $p_i$ does not access $\order(p_i, j+1)$ before
accessing $\order(p_i, j)$ at least once, but for every $j\prime \leq
j$, $p_i$ may access $\order(p_i, j\prime)$ several times before
accessing $\order(p_i, j+1)$ for the first time.  Since the memory is
anonymous, when a process accesses a register for the first time, say
register $\REG[x]$, we may map $x$ to any (physical) register that it
hasn't accessed yet.  However, when it accesses $\REG[x]$ again, it
must access the same register it has accessed before when referring to
$x$.\\

Let us now consider another division of the $k$ processes into sets.
We divide the $k$ processes into $k/ \delta$ disjoint sets, denoted
$Q_0,...,Q_{k/\delta -1}$, such that there are exactly $\delta$
processes in each set.  This partitioning  is achieved by assigning
process $p_i$ (where $i\in \{0,...,k-1\}$) to the set $Q_{\lfloor
  i/\delta\rfloor}$.  For example, when $k=6$ and $\delta =3$, $Q_0 =
\{p_0,p_1,p_2\}$, and $Q_1 = \{p_3,p_4,p_5\}$.  Again, such a  partitioning
is possible since $\ggcd(m,k)= \delta$ (bottom of
Figure~\ref{fig:proof-theorem-RMW-necessary}).

We notice that $Q_0$ includes the first process to take a step in the
execution $\rho$, in each one of the $\delta$ sets,
$P_0,...,P_{\delta -1}$.  Similarly, $Q_1$ includes the second process
to take a step in the execution $\rho$, in each one of the $\delta$
sets, $P_0,...,P_{\delta -1}$, and so on.

Since only comparisons for equality are allowed, and all registers are
initialized to the same value --which (to preserve anonymity) is not a
process identity-- in the execution $\rho$,
for each $i\in \{0,...,{n/\delta -1}\}$,
all the processes in the set $Q_i$ that
take the same number of steps must be at the same state. (This is
  because all the processes in  $Q_i$ are located at the
  same distance around the ring. At each lockstep, they invoke the
  Read/Modify/Write operation  into
  different locations, so because of the symmetry assumption,
  it is not possible to break the symmetry-between them,
(either all or none are elected.)  Thus, in the
run $\rho$, it is not possible to break symmetry within a set $Q_i$
($i\in \{0,...,{k/\delta -1}\}$), which implies that either all the
$\delta$ processes in the set $Q_i$ will be elected, or no
process in $Q_i$ will be elected.

Thus, the number of elected leaders in $\rho$ equals $\delta$ times
the number of $Q_i$ sets ($i\in \{0,...,{k/\delta -1}\}$) that all
their members were elected, and (by definition of $d$-election) it
must be a positive number.  That is, the number of elected leaders in
$\rho$ equals $a \delta$ for some integer $a\in\{1,..., k/\delta\}$.

Since in a $d$-election algorithm at most $d$ leaders are elected in
run $\rho$, it follows from the fact that for some positive integer
$a$, it must be the case that $a \delta \leq d$.  Thus, it must be the
case that $\ggcd(m,k)= \delta \leq d$.  Since $k$ was chosen
arbitrarily from $\{1,..., n\}$, it follows that a necessary
requirement for a symmetric $d$-election algorithm for $n\geq 2$
processes using $m$ anonymous RMW registers is that, for every $1\leq
k \leq n$, $\ggcd(m,k)\leq d$.
\renewcommand{\toto}{theorem-imposs-d-election}
\end{proofT}

\subsection{A \emph{d}-election algorithm  in RMW fully  anonymous systems}
\label{sec:d-election-algorithm}
\paragraph{Anonymous memory}
The  anonymous memory is made up  of $m$ RMW
registers $R[1..m]$, each initialized to the default value 0.
It is assumed that
$m\in  M(n,d)$ (recall that   $M(n,d) =\{k~:~ \forall ~\ell :
  1 < \ell\leq n\ : ~\gcd(\ell,k)\leq d\}$).

\paragraph{Local variables at each process $p_i$}
Each process $p_i$ manages the following set of local variables.
\begin{itemize}
\item $\counter_i$: used to store the number of RMW  registers {\it owned}
  by $p_i$. A process {\it owns} a register when  it is the last
  process that wrote a positive value into this register.
\item $\myview_i[1..n]$: array of Boolean values,
  each initialized to $\ffalse$. When
  $\myview_i[j]$ is equal to $\ttrue$,  $p_i$ owns the register  $R_i[j]$.
\item $\round_i$ (initialized to $0$): round number
  (rung number in the ladder metaphor, see below) currently
  attained by $p_i$ in its competition to be a leader.
  When $\round_i=n-d+1$, $p_i$ becomes a leader.
\item  $\competitors_i$: maximal number of
 processes that compete with $p_i$ when it executes a round.
\end{itemize}

\paragraph{Participation and output}
Any number of processes
can invoke the election algorithm.
A process exits the algorithm when it invokes  $\return({\tt res})$
where  ${\tt res}$ is  {\tt leader}  or  {\tt not leader}.

%

\paragraph{Description of the of the algorithm}
The code of each anonymous process $p_i$ appears in
Figure~\ref{fig:RMW:election}. When the process
$p_i$ invokes $\elect()$, it enters a repeat loop that
it will exit at line 11 if it is not elected, and at line
line 13 if it is elected.

Once in a new round,  $p_i$ first writes its new round number
in all the registers it owns, those are the registers $R_i[j]$
such that $\myview_i[j]=\ttrue$ (line~4). Then, it
strives to  own as many registers as possible (without compromising
liveness).
To this end, it considers all the registers $R_i[j]$
such that $R_i[j]<round_i$    (line~6). If such a register is equal to 0
(i.e., is not currently owned by another process), $p_i$
invoke $\compareandswap(R_i[j],0,round_i)$ to own it (line~7).
If it is the case, it accordingly increases $counter_i$  (line~8).

Then $p_i$ computes the maximal number of processes that, at round
$\round_i$, can compete with them (variable $\competitors_i$ at
line~9). There are then two cases.  If it owns fewer registers than the
average number $m/competitors$ (division on real numbers),
$p_i$ resets the registers it owns to their initial value (line~11),
and withdraws from the leader competition (line~12).
Otherwise, if $round_i <n-d+1$,
$p_i$ re-enters the repeat loop to progress to the next
round. If $round_i = n-d+1$, $p_i$ is one the at most $d$ leaders
(line~14).

Let us note that a (successful) assignment  of  a round number
 to  $R_i[j]$ by a process $p_i$ at line~7 has
  $R_i[j]=0$ as pre-condition and  $R_i[j]>0$ as post-condition.
 Moreover,  both the assignment of $R_i[j]$ at lines~4
and~11 have  $R_i[j]>0$ as pre-condition.
It follows that, between the lines~3 and~9
$counter_i$ counts the number of registers owned by $p_i$.

\begin{figure}[ht]
\small
\hrule
\vspace{0.4cm}
\centering
\noindent
\textsc{Algorithm 1: code of a  process $p_i$ in the fully anonymous RMW model
\\ (non mandatory participation)}%
\begin{tabbing}

\hspace{1.4em}\=\hspace{.4em}\=\hspace{1.3em}\=\hspace{1.5em}\=\hspace{1.5em}\=\hspace{1.5em}\=\kill

The initial value of all the shared registers is 0.\\
\\


{\bf operation} $\elect()$ {\bf is} \\
1 \> $\counter_i\leftarrow 0; \round_i\leftarrow 0$;
 \textbf{for each} $j\in \{1,...,m\}$ \textbf{do}
 $\myview_i[j] \leftarrow \ffalse$ \textbf{end\_do}\\

2 \> \textbf{repeat}\\



3 \>    \> $\round_i\leftarrow \round_i+1$
                                 \`// \texttt{progress to the next round}\\

4\>\>  \textbf{for each} $j\in \{1,...,m\}$ \textbf{do}
 \textbf{if} $\myview_i[j]$
\textbf{then} $R_i[j]\leftarrow \round_i$  \textbf{fi} \textbf{end\_do}
\`// {{\texttt{owned}}}\\ 

5\>\> \textbf{for each} $j\in \{1,...,m\}$  \textbf{do}
\`// \texttt{try to own more registers}\\

6\>\>\> \textbf{while} $R_i[j]< \round_i$  \textbf{do}
\`// \texttt{$\pp[j]< \round_i$ implies $\myview[j]=\ffalse$}\\

7\>\>\>\> $\myview_i[j]\leftarrow \compareandswap(R_i[j],0,\round_i)$
      \`// \texttt{try to own $R_i[j]$}\\
8\>\>\>\> \textbf{if} $\myview_i[j]$
      \textbf{then} $\counter_i\leftarrow \counter_i+1$ \textbf{fi}
      \textbf{end\_do} \textbf{end\_do} \`// \texttt{\small {own  $+ 1$}}\\

9\>\> $\competitors_i \leftarrow n-\round_i +1$
\`// \texttt{max \# of competing processes}\\ 

10\>\> \textbf{if} $\counter_i < m / \competitors_i$ \textbf{then}
\`// \texttt{too~many~competitors}\\

11\>\>\> \textbf{for each} $j\in \{1,...,m\}$  \textbf{do}
\textbf{if} $myview_i[j]$  \textbf{then} $R_i[j]\leftarrow 0$
\textbf{fi} \textbf{end\_do} \`// \texttt{free owned} \\

12\>\>\>  $\return$ ({\tt not~leader})  \textbf{fi}
 \`// \texttt{withdraw from the election}\\

13\> \textbf{until} $\round_i =n-d+1$  \textbf{end repeat} \\
14\> $\return$ ({\tt leader}).
\`// \texttt{$p_i$ is elected}
\end{tabbing}
\caption{\small{$d$-election for $n$
    anonymous processes and  $m\in M(n,d)$ anonymous RMW registers
\label{fig:RMW:election}}}
\vspace{0.3cm}
\hrule
\normalsize
\end{figure}

\subsection{Proof of Algorithm 1}
\label{sec:proof-election}
Let us say that ``process $p_i$ executes
round $r$'' when its local variable $round_i=r$.

\noindent
Reminder:
$m\in M(n,d)$ where $M(n,d) =\{k~:~ \forall ~\ell :
1 < \ell\leq n\ : ~\gcd(\ell,k)\leq d\}$.

%
%
%
%

\begin{lemma}
\label{lemma-invariant}
For every $r\in\{1,..., n-d+1\}$,
at most $n-r+1$ processes may execute round $r$.
In particular, at most $d$ processes may execute round $n-d+1$.
\end{lemma}

\begin{proofT}
The proof is by induction on the number of rounds.  The induction base
is simple since at most $n$ processes may execute round
$r=1$.  Let us assume (induction hypothesis) that the lemma holds
for round $r < n-d+1$ and prove that the lemma also holds
(induction step) for round $r+1$.  That is, we need to show that at most
$n-r$ processes execute round $r+1$.

Let $P_r$ be the set of processes that execute round $r$.
If $|P_r|< n-r+1$ then we are done, so let us assume that $|P_r|= n-r+1$.
Notice that, since $r < n-d+1$, $|P_r| > d$.

We have to show that at least one process in $P_r$ will not proceed to
round $r+1$, i.e., to show that at least one
process in $P_r$ will withdraw at line~12.   This amounts to show
that for at least one process $p_i\in P_r$ the predicate $\counter_i <
m/\competitors_i$ is evaluated to true when $p_i$ executes line~10
during round $r$.

Assume by contradiction that the predicate $\counter_i <
m/\competitors_i$ in line~13 is evaluated to false for each process
$p_i\in P_r$.  For each $1\leq i \leq |P_r|$, let $\counter(i)$
denotes the value of $\counter_i$ at that time (when the predicate is
evaluated to false).  Thus, for all $1\leq i \leq |P_r|$, $\counter(i)
\geq m/(n-r+1)$.
Hence, it follows from the following  facts,
\begin{enumerate}
\item
$\counter(1) + \cdots + counter(|P_r|)=m$,
\item
$\forall ~1\leq i \leq |P_r|~:~\counter(i) \geq m/(n-r+1)$, and
\item
$|P_r|= n-r+1$,
\end{enumerate}
 that
 $\forall ~1\leq i \leq |P_r|~:~\counter(i) = m/(n-r+1).$
 Moreover, as
\begin{enumerate}
\item
 $\counter(i)$ is a positive integer, we have
$\gcd(n-r+1,m) = n-r+1,$
%
\item
 $r< n-d+1$ it follows that $n-r+1 > d,$
\end{enumerate}
from which  follows that $\gcd(n-r+1,m) > d$, which  contradicts the
  assumption that $m\in M(n,d)$.
\renewcommand{\toto}{lemma-invariant}
\end{proofT}


\begin{lemma}
\label{lemma-election-safety}
At most $d$ processes are elected.
\end{lemma}

\begin{proofL}
  The proof is an immediate consequence of Lemma~\ref{lemma-invariant},
  which states that at most $d$ processes may execute round $n-d+1$.
  If they do not withdraw from the competition, each of these processes
  exits the algorithm at line~14, and considers it is a leader.
\renewcommand{\toto}{lemma-election-safety}
\end{proofL}


\begin{lemma}
\label{lemma-election-liveness}
For every $r\in\{1,..., n-d+1\}$,
at least one process executes round $r+1$.
In particular, at least one process executes round $n-d+1$
 at the end of which it claims it is a  leader.
\end{lemma}

\begin{proofL}
  \MMR{ Considering the (worst) case where the $n$ processes execute
    round $r=1$, we show that at least one process attains
    round~$2$.  To this end, let us assume by contradiction that no
    process attains round~2. This means that all the processes
    executed line 10 and found the predicate equal to true (they all
    withdrew) hence each process $p_i$ is such that $counter_i <
    m/(n-r+1) =m/n$.  Using the notations and the observations of
    Lemma~\ref{lemma-invariant}, we have
\begin{enumerate}
\item
$|P_r|= n$,
\item
$\counter(1) + \cdots + counter(n)=m$,
\item
$\forall ~1\leq i \leq n~:~\counter(i) < m/(n-r+1)= m/n$.
\end{enumerate}
If then follows from the last item that $\counter(1) + \cdots +
counter(n) < n \times m/n=m$ which contradicts the second item.  It
follows from this contradiction that there is at least one process for
which the predicate of line~10 is false at the end of round 1, and
consequently this process progresses to round $r=2$.

Assuming now by induction that at most $(n-r+1)$ processes execute
round $r$, we show that at least one process progresses to round
$r+1$.  The proof follows from the three previous items where $|P_r|=
n-r+1$ (item 1), $\counter(1) + \cdots + counter(|P_r|)=m$ (item 2),
and $\forall ~1\leq i \leq |P_r|~:~\counter(i) < m/(n-r+1)$ (item 3),
from which we conclude $\counter(1) + \cdots + counter( |P_r|) < n
\times (m-r+1) (m/ (m-r+1))$, i.e. $m<m$, a contradiction.  It follows
that at least one process executes the round $r+1$ during which it
finds the predicate of line 10 false and consequently progresses to
the next round if $r<n-d+1$.  If $r=n-d+1$, the process executes
line~14 and becomes a leader.  }
  \renewcommand{\toto}{lemma-election-liveness}
\end{proofL}


\begin{theorem}
\label{theorem-algorithm-1}
Let $m$, $n$ and $d$ be such that $m\in M(n,d)$,
and assume at least one process invokes $\elect()$.  Algorithm~{\em 1}
(Fig.~{\em\ref{fig:RMW:election}}) solves $d$-election
in a fully anonymous system where communication is through {\em RMW}
registers.
\end{theorem}

\begin{proofT}
  The proof follows directly from the
  lemmas~\ref{lemma-election-safety}
  and~\ref{lemma-election-liveness}.
  \renewcommand{\toto}{theorem-algorithm-1}
\end{proofT}

\section{Exact \emph{d}-election in the RMW Model \\
  Where  Participation is Required}
\label{sec-exact-d-election}

This section considers the fully anonymous RMW model in which all the
processes are required to participate.  In such a context, it presents
an exact $d$-election algorithm that assumes that $d$ is a multiple of
$\ggcd(m,n)$. It also shows that this condition is necessary for exact
$d$-election in such a system model.

\subsection{A necessary condition for  exact \emph{d}-election}

The following theorem, which considers
anonymous memory and non-anonymous processes with the symmetry constraint,
has been stated and proved in~\cite{GIRT20-podc}.
\begin{theorem}[\textbf{See \cite{GIRT20-podc}}]
\label{theorem-imposs-exact-d-election}
There is no symmetric exact $d$-election algorithm in the {\em RMW}
communication model for $n\geq 2$ processes using $m$ anonymous
registers if $\ggcd(m,n)$ does not divide~$d$.
\end{theorem}
As in Section \ref{imposs-d-election-anonymous-memory} this
impossibility still holds in the RMW model where both the processes and
the memory are anonymous, and in the model where communication is
through anonymous RW registers.

\subsection{An exact \emph{d}-election algorithm}

\paragraph{Anonymous memory}
All the registers of the anonymous memory $R[1..m]$ are RMW
registers initialized to $0$. Moreover, the size $m$ of the
memory is such that $\ggcd(m,n)$ divides $d$.

An anonymous register $R[x]$ will successively contain the values 1,
2, ...  where the increases by 1 are produced by  successful
executions of $\compareandswap(R[x],val, val+1)$ issued by the
processes (lines~6 and~8 in Figure~\ref{fig:RMW:exact-d-election}).
The fact that a process can increase the value of a
register to $val+1$ only if its current value is $val$ is the key of
the algorithm.

\paragraph{Underlying principle}
The key idea that governs the algorithm is Bezout's identity, a
Diophantine equation that relates any pair of positive integers
according to their Greatest Common Divisor\footnote{This principle has
  already been used in~\cite{GIRT20-podc} to solve exact $d$-election
  with a {\it symmetric} algorithm in a system where the
  (non-anonymous) processes cooperate through an anonymous RW
  registers.}.

\begin{theorem}[Bezout, 1730-1783]
Let $m$ and $n$ be two positive integers and let $d=\ggcd(m,n)$.
There are two positive  integers $u$ and $v$ such that $u\times m= v\times n
+d$.\footnote{The pair
  $\langle u,v\rangle$ is not unique. Euclid's $\ggcd(m,n)$ algorithm
  can be used to compute such pairs.}
\end{theorem}

\sloppy  Consider a rectangle made up of $u\times m$ squares.  On one side,
this means that $u$ squares are associated with each of the $m$
anonymous registers. On another side, each of the $n$ processes
progresses until it has ``captured'' $v$ squares (from an operational
point of view, the capture of a square is a successful
invocation of  $\compareandswap(R[x],val,val+1)$.

Then, when $v\times n$ squares have been captured by the processes,
 each process
competes to capture one more square.  As it remains only $d= u\times m
- v\times n$ squares, the processes that succeed in capturing one more
square are the $d$ leaders.

\paragraph{Local variables at each process $p_i$}
\begin{itemize}
\item $\won_i$ (initialized to $0$):   number of squares
 captured by $p_i$.
\item $\ssum_i$ (initialized to $0$):  local view of the numbers
  of squares captured by all the processes.
\item $\myview_i[1..m]$: local copy (non-atomically obtained) of the
  anonymous memory $R[1..m]$.
\end{itemize}

\begin{figure}[ht]
\small
\hrule
\vspace{0.4cm}
\centering
\noindent
\textsc{Algorithm 2: code of a process $p_i$ in the fully anonymous RMW model}\\

\textsc{(mandatory participation)}
\begin{tabbing}
\hspace{1.5em}\=\hspace{1.5em}\=\hspace{6em}\=\kill
$u$ and $v$: smallest positive integers such that $u\times m= v\times n +d$\\
The initial value of all the shared registers is 0.\\
~\\

%

\hspace{1.5em}\=\hspace{.5em}\=\hspace{1.em}\=\hspace{1.2em}\=\hspace{1.5em}\=\hspace{1.5em}\=\kill

{\bf operation} $\elect()$ {\bf is} \\

1 \> \textbf{repeat}\\

2 \>\>\textbf{for each} $j\in \{1,...,m\}$ \textbf{do}
     $\myview_i[j] \leftarrow R[j]$ \textbf{end\_do}
            \`// \texttt{read of the anony  mem.}\\
3 \>\>
$\ssum_i \leftarrow \myview_i[1] + \cdots + \myview_i[m]$
 \`// \texttt{\# successful $\compareandswap()$ seen}\\

4 \>\> \textbf{if} $\exists~ x\in \{1,...,m\}~:~ \myview_i[x]<u$ \textbf{then}\\

5 \>\>\> \textbf{if} $\won_i <v$ \textbf{then} \\ 

6 \>\>\>\> \textbf{if} $\compareandswap(R[x],\myview_i[x],\myview_i[x]+1)$
  \textbf{then} $\won_i\leftarrow \won_i+1$  \textbf{fi} \textbf{fi}\\

7  \>\>\> \textbf{if} $\ssum_i \geq v\times n$ \textbf{then}
       \`// \texttt{$\ssum_i \geq v\times n$ implies  $\won_i=v$} \\

8 \>\>\>\> \textbf{if} $\compareandswap(R[x],\myview_i[x],\myview_i[x]+1)$
  \textbf{then} $\return$ ({\tt leader})  \textbf{fi} \textbf{fi}\\

9 \>\> \textbf{fi} \\

10\> \textbf{until} $\ssum_i = u\times m$  \textbf{end repeat}\\
11\> $\return$ ({\tt not~leader}).
\end{tabbing}
\caption{
  Exact $d$-election for $n$
    anonymous processes and $m$ RMW anonymous registers}
\label{fig:RMW:exact-d-election}
\vspace{0.3cm}
\hrule
\normalsize
\end{figure}

\paragraph{Description of the algorithm}
Assuming $d$ is a multiple of $\ggcd(m,n)$ and all the processes
participate, Algorithm 2 (described in
Figure~\ref{fig:RMW:exact-d-election}) solves exact $d$-election
for $n$ anonymous processes and $m$ RMW anonymous registers.

When it invokes $\elect()$, a process $p_i$ enters a repeat loop
lines~1-10. Each time it enters the loop, $p_i$ asynchronously reads
the anonymous memory non-atomically (line~2) and then counts in $\ssum_i$
the number of squares
that  have been captured by all  processes as indicated by the previous
 asynchronously reads (line~3).

If $p_i$ sees a register $R[x]$ that has been captured less than $u$ times
(line 4), there are two cases.
\begin{itemize}
\item If $\won_i<v$, $p_i$ tries to capture one of the $u$
 squares of $R[x]$.  To
  this end $p_i$ uses the RMW operation: it invokes
  $\compareandswap(R[x],\myview_i[x],\myview_i[x]+1)$.  If it is
  successful, it increases $\won_i$, the number of squares it has
  captured so far (line~6).

\item If $\ssum_i\geq  v\times n$ (we have then  $\won_i=v$),
  $p_i$ strives to capture one more square (line~8).
  If it is successful, it is elected as  of the $d$ leaders.

  In the other case, if  $\ssum_i= u\times m$, all the squares have been
  captured, so $p_i$ is not a leader (line~11).
  Otherwise, $p_i$ re-enters the repeat loop.
\end{itemize}

\noindent
{\sf Remark}
  Let $\alpha$ and $\beta$ be two integers such that
  $m= \alpha\times \ggcd(m,n)$ and  $n= \beta\times \ggcd(m,n)$.
  The equations $u\times m= v\times n +d$ and $d= \ell \times \ggcd(m,n)$
  give rise to the equation $u\times \alpha = v \times \beta  +\ell$,
  which can be used to  obtain a more efficient version of the algorithm.

\subsection{Proof of Algorithm 2}

\begin{theorem}
\label{theorem-algorithm-2}
Let $m$, $n$ and $d$ be such that $\ggcd(m,n)$ divides $d$, and assume
all the processes invoke $\elect()$.  Algorithm~{\em 2}
(Fig.~{\em\ref{fig:RMW:exact-d-election}}) solves exact $d$-election
in a fully anonymous system where communication is through {\em RMW}
registers.
\end{theorem}

\begin{proofT}
  Let us first observe that, due to the atomicity of
  $\compareandswap()$, if several processes
  invoke  $\compareandswap(X,v,v+1)$ on the very same register $X$ whose
  value is $v$, exactly one of of them succeeds in writing $v+1$.
  It follows that each of the  $u\times m$ squares is captured by
  only one process. Moreover, due to the predicate of line~5, each
  process eventually captures $v$ squares.
  Once this occurs, it remains $d$ squares, which are captured by
  $d$ distinct processes at line~8 (these processes are distinct because,
  once a process captured such a square, it returns the value
  $\tt leader$ and stops executing).
  Moreover,  a process can capture one of the $d$
  remaining squares only after each process
  has captured $v$ squares at line~6.
  It follows that exactly $d$ processes exit the algorithm at line~7
  with a successful $\compareandswap()$, and the $(n-d)$ other processes
  exit the algorithm at line~11.
\renewcommand{\toto}{theorem-algorithm-2}
\end{proofT}

\section{\emph{d}-Election in the RMW Model Where Participation is Required}
\label{sec-d-election-required}

This section considers the fully anonymous RMW model
in which all the processes are required to
participate.  In such a context, it presents a $d$-election
algorithm where
$\ggcd(m,n)\leq d$. It also shows that this condition is necessary for
$d$-election in such a system model.

\subsection{A necessary condition for \emph{d}-election}
The following theorem, which considers
anonymous memory and non-anonymous processes with the symmetry constraint,
has been stated and proved in~\cite{GIRT20-podc}.
\begin{theorem}[\textbf{See \cite{GIRT20-podc}}]
\label{theorem-imposs-d-election-required}
There is no symmetric $d$-election algorithm in the {\em RMW}
communication model for $n\geq 2$ processes using $m$ anonymous
registers if $\ggcd(m,n) > d$.
\end{theorem}
Clearly, this
impossibility still holds in the RMW model where both the processes and
the memory are anonymous, and in the model where communication is
through anonymous RW registers.

\subsection{A necessary and sufficient condition for \emph{d}-election}

The following corollary is an immediate consequence of
Theorem~\ref{theorem-algorithm-2}.
\begin{corollary}
\label{corollary:gcd}
For any pair $\langle n, m\rangle$, it is always possible to solve
exact $\gcd(n,m)$-election in a fully anonymous system where
communication is through {\em RMW} registers.
\end{corollary}
Let us also observe that any exact  $d$-election algorithm
trivially solves $d$-election (but then the bound
is then not tight). We also have the following theorem.
\begin{theorem}
\label{theorem-algorithm-3}
For any pair  $\langle n, m\rangle$, it is possible to solve
$d$-election in a fully anonymous system where
communication is through {\em RMW} registers if and only if $\ggcd(n,m) \leq d$.
\end{theorem}

\begin{proofT}
 If direction.
 For any pair $\langle n, m\rangle$ such that $\ggcd(n,m) \leq d$, it
 is possible to solve $d$-election by running an exact
 $\ggcd(n,m)$-election algorithm, which exists due to
 Corollary~\ref{corollary:gcd}.

As the fully anonymous model is a weaker model than the symmetric
model, the ``Only if'' direction follows from
Theorem~\ref{theorem-imposs-d-election-required}.
\renewcommand{\toto}{theorem-algorithm-3}
\end{proofT}

\section{Impossibility in the RW Communication Model}
\label{sec-RW-impossibility}
\begin{theorem}
\label{theo--imposs-RW-model}
There is neither   $d$-election nor exact $d$-election algorithms
in the process  anonymous {\em RW} non-anonymous communication model.
\end{theorem}

\begin{proofT}
  Assuming such an algorithm exists, let us order the participating
  processes
  in some fixed order, e.g., $p_1$, ..., $p_x$
  ($x=n$ in the case where full  participation is required).
Let us consider  in such a setting a lock-step execution in which $p_1$
  executes its first (read or write operation on a shared register)
  operation, then $p_2$ executes its first operation, etc., until
  $p_x$ that executes its first (read or write)  operation
  on the shared non-anonymous memory. As all processes have the
  same code, they all execute the same operation on the same register
  and are consequently in the same local state after having executed
  their first operation.  The same occurs after they have their (same)
  second operation, etc.  It follows that, whatever the number of
  steps executed in a lock-step manner by the processes, they all are
  in the same local state. So, it is impossible to break their
  anonymity (that would allow us to elect some of them).
\renewcommand{\toto}{theo--imposs-RW-model}
\end{proofT}

Let us consider an anonymous memory in which the memory adversary
associates the same address mapping to all the processes (i.e.,
$\forall i,j \in \{1,\cdots,n\}$ and $x\in\{1,\cdots,m\}$
we have $f_i (x)= f_j(x)$, see Section~\ref{def:anonymous-memory}).
In this case,
the model boils down to the process anonymous and non-anonymous
memory.  The next corollary is then an immediate consequence of the
previous theorem.
\begin{corollary}
\label{coro--imposs-RW-model}
There is neither  $d$-election nor exact $d$-election algorithms
in the fully  anonymous {\em  RW} communication model.
\end{corollary}

\section{Conclusion}
This article has investigated the $d$-election problem in fully
anonymous shared memory systems. Namely, systems where not only the
processes are anonymous but the shared memory also is anonymous in the
sense that there is no global agreement on the names of the shared
registers (any register can have different names for distinct
processes).  Assuming RMW atomic registers, it has shown that both the
$d$-election problem (at least one and at most processes are elected)
and the exact $d$-election problem (exactly $d$ processes are elected)
can be solved in such an adversarial context if and only if the three
model parameters $n$ (number of processes), $m$ (size of the anonymous
memory), and $d$ (number of leaders) satisfy some properties.
These necessary and sufficient conditions are:
\begin{itemize}
\item
$m\in M(n,d)$
for solving $d$-election when participation is not required,
\item
$\ggcd(m,n)$ divides $d$
for solving exact $d$-election when participation is required, and
\item
$\ggcd(m,n)\leq d$
for solving $d$-election when participation is required.
\end{itemize}
It has also been shown that,
\begin{itemize}
\item
neither $d$-election nor exact $d$-election can be solved
in a fully anonymous system where communication is through
atomic RW registers.
\end{itemize}
%
  This work complements previously known research on  the symmetry-breaking
  problem (election) in the context of  fully anonymous RW/RMW systems.
  A very challenging problem remains to be solved: are there
  other non-trivial functions that can be solved in the fully  anonymous RW/RMW
  setting.

  \section*{Acknowledgments}
 The authors want to thank the referees for their cosntructive comments.

\appendix
\section{The case where $m=1$}
When the anonymous memory is made up of a single register $R$,
we have $\gcd(1,n)=1\leq 1$.  In this case there is  a very simple
$d$-election algorithm described below, where the single anonymous register
is initialized to $0$.

\begin{figure}[ht]
\small
\hrule
\vspace{0.4cm}
\centering
\noindent
\textsc{Algorithm 3: code of a process $p_i$ when $m=1$}\\

\textsc{(participation required or not required)}
\begin{tabbing}
\hspace{1.5em}\=\hspace{.5em}\=\hspace{1.5em}\=\hspace{1.5em}\=\hspace{1.5em}\=\hspace{1.5em}\=\kill

{\bf operation} $\elect()$ {\bf is} \\

1 \> \textbf{repeat forever}\\

2 \>\>   $\myview_i \leftarrow R$
         \`// \texttt{atomic read the anonymous  memory}\\
3
\>\> \textbf{if} $myview_i\geq d$ \textbf{then} $\return$ ({\tt not~leader})
     \textbf{fi}\\
4
\>\> \textbf{if} $\compareandswap(R,\myview_i,\myview_i+1)$
           \textbf{then} $\return$ ({\tt leader}) \textbf{fi}\\

5\>  \textbf{end repeat}.
\end{tabbing}
\caption{
  $d$-election for $n$
    anonym. processes when the anonym.  memory is a single RMW register}
\label{fig:m=1RMW:exact-d-election}
\vspace{0.3cm}
\hrule
\normalsize
\end{figure}


\end{document}